\documentclass[3p,times,twocolumn]{elsarticle}
 \biboptions{comma,sort&compress}
 
\usepackage{graphicx}
\usepackage{amsmath}
\usepackage{here}
\usepackage{ecrc}
\volume{00}
\firstpage{1}
\journalname{Nuclear and Particle Physics Proceedings}

\jid{nppp}
\jnltitlelogo{Nuclear and Particle Physics Proceedings}

\usepackage{amssymb}
\usepackage[colorlinks=true,citecolor=blue,anchorcolor=red,menucolor=red, linkcolor=red,filecolor=red,runcolor=red,urlcolor=blue,frenchlinks=true]{hyperref}
\usepackage[figuresright]{rotating}

\usepackage{inputenc}
\usepackage{graphics}
\usepackage{dsfont}
\usepackage{mathtools}
\usepackage{autobreak}
\usepackage{diagbox}
\usepackage{multirow}
\usepackage{longtable}
\usepackage{txfonts}
\usepackage{bm}
\usepackage{amsfonts}
\usepackage{subfigure}

\usepackage{overpic}
\usepackage{indentfirst}
\usepackage{cancel} 
\usepackage{slashed}  
\usepackage{cases}
\usepackage{color}
\usepackage{natbib}

\newcommand{\GGb}{\langle \alpha_s GG\rangle}

\newcommand{\GGGb}{\langle g_s^3G^3\rangle}

\begin{document}

\begin{frontmatter}

\title{Towards heavy double-gluon hybrid mesons with exotic quantum numbers in QCD sum rules
} 
 \author[au-a]{Ding-Kun Lian\corref{cor1}}
 \cortext[cor1]{Corresponding author.}
 \ead{liandk@mail2.sysu.edu.cn}
 \author[au-a]{Qi-Nan Wang}
 \author[au-a]{Xu-Liang Chen}
 \author[au-a]{Peng-Fei Yang}
 \author[au-a,au-b]{Wei Chen}
 \ead{chenwei29@mail.sysu.edu.cn}
 \author[au-c]{Hua-Xing Chen}
 \address[au-a]{School of Physics, Sun Yat-Sen University, Guangzhou 510275, China}
 \address[au-b]{Southern Center for Nuclear-Science Theory (SCNT), Institute of Modern Physics, Chinese Academy of Sciences, Huizhou 516000, Guangdong Province, China}
 \address[au-c]{School of Physics, Southeast University, Nanjing 210094, China}

\pagestyle{myheadings}
\markright{ }
\begin{abstract}
\noindent
The double-gluon hybrid meson configuration was recently proposed and investigated within QCD sum rules. In this talk, we discuss the color structures of the double-gluon hybrid meson and construct current operators with exotic quantum numbers $J^{PC}=1^{-+}$ and $2^{+-}$ for two of the structures. In the framework of QCD sum rules, we consider the condensates up to dimension-8 at the leading order of $\alpha_{s}$ for both charmonium and the bottomonium systems. The results indicate that the masses of the $1^{-+}$ and $2^{+-}$ charmonium double-gluon hybrid mesons are approximately $6.1-7.2$ GeV and $6.3-6.4$ GeV, respectively. As for the bottomonium systems, their masses fall within the range of $13.7-14.3$ GeV and $12.6-13.3$ GeV for the $1^{-+}$ and $2^{+-}$ channels, respectively. Additionally, the charmonium hybrids could be produced in the radiative decays of bottomonium mesons in BelleII experiment.
\end{abstract}
\end{frontmatter}
\section{Introduction}
In the conventional quark model~\cite{ParticleDataGroup:2022pth,Gell-Mann:1964ewy,Zweig:570209}, hadrons have two types: $q\bar{q}$ mesons and $qqq$ baryons. However, quantum chromodynamics (QCD) allows the existence of the exotic hadrons beyond ordinary mesons and baryons, such as multiquark states, hybrid mesons and glueballs, etc. A typical hybrid meson is composed of a quark-antiquark pair and one valence gluon ($\bar{q}Gq$). Due to the existence of gluonic degree of freedom, hybrid mesons can carry exotic quantum numbers ($J^{PC}=0^{--},even^{+-},odd^{-+}$), which are forbidden in the ordinary $q\bar{q}$ systems. Taking a step further, the double-gluon hybrid meson, composed of a quark-antiquark pair and two valence gluons, was recently proposed and studied in the light quark sector~\cite{Chen:2021smz,Su:2022fqr,Su:2023jxb} and subsequently extended to the heavy quark sector~\cite{Tang:2021zti,Su:2023aif}. There is no doubt that the addition of one more valence gluon field will enrich the spectra of hybrid mesons. In this talk, we shall focus on the heavy quarkonium double-gluon hybrid mesons with exotic quantum numbers $J^{PC}=1^{-+}$ and $2^{+-}$ in the framework of QCD sum rule~\cite{Shifman:1978bx,Reinders1985,Colangelo2000,Narison2007,Gubler:2018ctz}.

The rest of this paper is organized as follows. In section~\ref{sec2}, we shall discuss the color structures of the double-gluon hybrid operator and construct the interpolating currents carrying $J^{PC}=1^{-+}, 2^{+-}$. In section~\ref{sec3}, we apply the QCD sum rule method to study the current operators. In section~\ref{sec4}, we perform  numerical analyses and extract hadron masses of the charmonium and bottomonium double-gluon hybrids. The last section is a brief summary.

\section{Double-gluon hybrid meson interpolating currents}\label{sec2}
A double-gluon hybrid meson operator $\bar{Q}GGQ$ is composed of a quark-antiquark $\bar{Q}Q$ field and a two-gluon glueball $GG$ field. In the color SU(3) symmetry, the color structures for the $\bar{Q}GGQ$ operator can be obtained as
\begin{equation}
\begin{aligned}
&(\overline{\mathbf{3}} \otimes \mathbf{3})_{[\bar{Q}Q]} \otimes(\mathbf{8} \otimes \mathbf{8})_{[GG]}\\ =&(\mathbf{1} \oplus \mathbf{8})_{[\bar{Q}Q]} \otimes(\mathbf{1} \oplus \mathbf{8} \oplus \mathbf{8} \oplus \mathbf{10} \oplus \overline{\mathbf{1 0}} \oplus \mathbf{27})_{[GG]} \\
=&(\mathbf{1} \otimes \mathbf{1}) \oplus(\mathbf{8} \otimes \mathbf{8})\oplus(\mathbf{8} \otimes \mathbf{8})\oplus\cdots\,,
\end{aligned}
\end{equation}
in which the color singlet structure may come from the $(\mathbf{1}_{[\bar{Q}Q]} \otimes \mathbf{1}_{[GG]})$ and $(\mathbf{8}_{[\bar{Q}Q]} \otimes \mathbf{8}_{[GG]})$ terms. Neglecting the Lorentz structure, a color singlet double-gluon hybrid meson operator has the following form
\begin{align}\label{eq: operatorform1}
    \bar{Q}_{a}(T^{s}T^{t})_{ab}Q_{b}G^{s}G^{t}\, ,
\end{align}
where $Q$ represents a quark field, $G^{s}$ is the gluon field strength, $a, b$ are the color indices, $T^{s}=\lambda^s/2$ ($s=1, 2,...,8$) are the generators of SU(3) group. The generators satisfy the following relation
\begin{align}\label{Trelation}
    T^{s}T^{t}=\frac{1}{2}\left[\frac{1}{3}\delta^{st}\mathds{1}_{3\times3} +(d^{rst}+i f^{rst})T^{r}\right]\, ,
\end{align}
where $\mathds{1}_{3\times3}$ is the three-dimension identity matrix, $f^{rst}$ and $d^{rst}$ are the completely antisymmetric structure constants and completely symmetric constants of SU(3) group respectively. With this relation, the double-gluon hybrid meson operator in Eq.~\eqref{eq: operatorform1} can be separated into three terms with different color structures
\begin{equation}\label{eq: operatorform2}
  \begin{aligned}
    &\bar{Q}_{a}(T^{s}T^{t})_{ab}Q_{b}G^{s}G^{t}\\
    =&\frac{1}{6}\bar{Q}_{a}Q_{a}G^{t}G^{t}+\frac{1}{2}\bar{Q}_{a}(T^{r})_{ab}Q_{b}d^{rst}G^{s}G^{t}\\
    &+\frac{i}{2}\bar{Q}_{a}(T^{r})_{ab}Q_{b}f^{rst}G^{s}G^{t}\, .
  \end{aligned}
\end{equation}
In the above decomposition, the first term corresponds to the color structure $(\mathbf{1}_{[\bar{Q}Q]} \otimes \mathbf{1}_{[GG]})$ which denotes the meson-glueball molecule state~\cite{Petrov:2022ipv}. The second and third terms are in the octet-octet $(\mathbf{8}_{[\bar{Q}Q]} \otimes \mathbf{8}_{[GG]})$ color structure,  containing symmetric and antisymmetric two-gluon glueball operators, respectively. We shall use the last two color structures to study the heavy quarkonium double-gluon hybrid mesons, i.e. charmonium system and bottomonium system.  
\begin{table}[htbp]
    \centering
    \renewcommand\arraystretch{1.5}
    \caption{Parity and {\it C}-parity of the color octet quark-antiquark and double-gluon operators.}
    \setlength{\tabcolsep}{3mm}
    {
    \begin{tabular}{ccc}
        \hline\hline
     Operator   &  $P$   &$C$\\ \hline
        $\bar{Q}_{a}Q_{b}$ &  $+$   &   $+$\\
        $\bar{Q}_{a}\gamma_{5}Q_{b}$   &  $-$   &$+$\\
        $\bar{Q}_{a}\gamma_{\mu}Q_{b}$   &  $(-1)^{\mu}$   &$-$\\
        $\bar{Q}_{a}\gamma_{\mu}\gamma_{5}Q_{b}$   &  $-(-1)^{\mu}$   &$+$\\
        $\bar{Q}_{a}\sigma_{\mu\nu}Q_{b}$   &  $(-1)^{\mu}(-1)^{\nu}$   &$-$\\
                $d^{rst}G^{s}_{\alpha\beta}G^{t}_{\gamma\delta}T^{r}$ &  $(-1)^{\alpha}(-1)^{\beta}(-1)^{\gamma}(-1)^{\delta}$   &   $+$\\
        $f^{rst}G^{s}_{\alpha\beta}G^{t}_{\gamma\delta}T^{r}$ &  $(-1)^{\alpha}(-1)^{\beta}(-1)^{\gamma}(-1)^{\delta}$   &   $-$\\
        $d^{rst}\widetilde{G}^{s}_{\alpha\beta}G^{t}_{\gamma\delta}T^{r}$ &  $-(-1)^{\alpha}(-1)^{\beta}(-1)^{\gamma}(-1)^{\delta}$   &   $+$\\
        $f^{rst}\widetilde{G}^{s}_{\alpha\beta}G^{t}_{\gamma\delta}T^{r}$ &  $-(-1)^{\alpha}(-1)^{\beta}(-1)^{\gamma}(-1)^{\delta}$   &   $-$\\
        \hline \hline
    \end{tabular}
    }
    
\label{table:PC_qq+gg}
\end{table}

Now we need to insert the Lorentz structures into the double-gluon hybrid meson operators to determine their $J^{PC}$ quantum numbers. 
We summarize the parities and {\it C}-parities of the color octet quark-antiquark $\bar{Q}Q$ and two-gluon $GG$ operators with various Lorentz structures in Table~\ref{table:PC_qq+gg}. Combining these $\bar{Q}Q$ and $GG$ operators, we can construct the interpolating currents for the heavy quarkonium double-gluon hybrid mesons with exotic quantum numbers $J^{PC}=1^{-+}$ and $2^{+-}$ as the following. 

We can build six double-gluon hybrid meson currents with one Lorentz index carrying exotic quantum numbers $J^{PC}=1^{-+}$:
\begin{equation}\label{eq: currentoperator1}
    \begin{aligned}
        J_{\mu}^{1}&=\bar{Q}_{a}T^{r}_{ab}\gamma_{\rho}Q_{b}g_s^2f^{rst}G^{s}_{\mu\nu}G^{t,\,\nu\rho}\,,\\
            J_{\mu}^{2}&=\bar{Q}_{a}T^{r}_{ab}\gamma_{\rho}Q_{b}g_s^2f^{rst}\widetilde{G}^{s}_{\mu\nu}\widetilde{G}^{t,\,\nu\rho}\,, \\
            J_{\mu}^{3}&=\bar{Q}_{a}T^{r}_{ab}\gamma_{\mu}Q_{b}g_s^2f^{rst}G^{s}_{\alpha\beta}G^{t,\,\alpha\beta}\,,\\
        J_{\mu}^{4}&=\bar{Q}_{a}T^{r}_{ab}\gamma_{\mu}\gamma_{5}Q_{b}g_s^2d^{rst}\widetilde{G}^{s}_{\alpha\beta}G^{t,\,\alpha\beta}\,, \\
        J_{\mu}^{5}&=\bar{Q}_{a}T^{r}_{ab}\gamma_{\rho}\gamma_{5}Q_{b}g_s^2d^{rst}\widetilde{G}^{s}_{\mu\nu}G^{t,\,\nu\rho}\,, \\
        J_{\mu}^{6}&=\bar{Q}_{a}T^{r}_{ab}\gamma_{\rho}\gamma_{5}Q_{b}g_s^2d^{rst}G^{s}_{\mu\nu}\widetilde{G}^{t,\,\nu\rho}\, ,
    \end{aligned}
\end{equation} 
where $g_s$ is the strong coupling and $\widetilde{G}^{s}_{\mu\nu}=\frac{1}{2}\varepsilon _{\mu\nu\alpha\beta}G^{s,\,\alpha\beta}$ is the dual of gluon field strength. As demonstrated in ~\ref{appendix: relation}, these six currents are not independent because of the symmetries of the glueball operators. Noting that the currents $J_{\mu}^{1}$ and $J_{\mu}^{2}$ were first constructed and studied in Ref.~\cite{Tang:2021zti}, in which their quantum numbers were considered incorrectly as $J^{PC}=1^{--}$. Moreover, the calculations in Ref.~\cite{Tang:2021zti} are in contradiction to the relation $J_{\mu}^{1}=-J_{\mu}^{2}$ in Eq.~\eqref{eq: relation1-+}.

For the $2^{+-}$ sate, only one double-gluon hybrid meson current with two Lorentz indices can be constructed:
\begin{equation}\label{eq: currentoperator2}
    \begin{aligned}
    J_{\mu\nu}^{1}&=\bar{Q}_{a}T^{r}_{ab}\gamma_{5}Q_{b}g_s^2f^{rst}\widetilde{G}^{s}_{\mu\alpha}G^{t,\,\alpha}_{\nu}\, .
    \end{aligned}
\end{equation}
However, the two-point correlation function induced by $J_{\mu\nu}^{1}$ at the leading order of $\alpha_s$ only contributes to the perturbation theory. The OPE series of the current $J_{\mu\nu}^{1}$ are too simple to study the $2^{+-}$ sate. 

In order to investigate the $2^{+-}$ double-gluon hybrid mesons, we have to use the interpolating currents with more Lorentz indices. In Refs~\cite{Chen:2021smz,Su:2022fqr,Su:2023jxb,Su:2023aif}, the authors adopted the currents with four Lorentz indices by symmetrizing Lorentz indices to extract the $2^{+-}$ state, but the currents became more complicated.
Here we use the currents with three Lorentz indices to study the $2^{+-}$ state, as our previous work for the one-gluon hybrid mesons in Ref.~\cite{Wang:2023whb}. There are four double-gluon hybrid meson interpolating currents with three Lorentz indices for the $2^{+-}$ state:
\begin{equation}\label{eq:currentoperator3}
    \begin{aligned}
        J_{\mu\nu\rho}^{1}&=\bar{Q}_{a}T^{r}_{ab}\gamma_{\alpha}Q_{b}g_s^2d^{rst}G^{s}_{\mu\nu}G^{t,\,\alpha}_{\rho}\,,\\
            J_{\mu\nu\rho}^{2}&=\bar{Q}_{a}T^{r}_{ab}\gamma_{\alpha}Q_{b}g_s^2d^{rst}G^{s}_{\mu\nu}\widetilde{G}^{t,\,\alpha}_{\rho}\,,\\
        J_{\mu\nu\rho}^{3}&=\bar{Q}_{a}T^{r}_{ab}\gamma_{\alpha}\gamma_{5}Q_{b}g_s^2f^{rst}G^{s}_{\mu\nu}G^{t,\,\alpha}_{\rho}\,,\\
        J_{\mu\nu\rho}^{4}&=\bar{Q}_{a}T^{r}_{ab}\gamma_{\alpha}\gamma_{5}Q_{b}g_s^2f^{rst}G^{s}_{\mu\nu}\widetilde{G}^{t,\,\alpha}_{\rho}\, .
    \end{aligned}
\end{equation}
One can replace $G^{s}_{\mu\nu}$ with $\widetilde{G}^{s}_{\mu\nu}$ in the above equations to obtain the dual currents $\tilde{J}_{\mu\nu\rho}^{1}$, $\tilde{J}_{\mu\nu\rho}^{2}$, $\tilde{J}_{\mu\nu\rho}^{3}$, $\tilde{J}_{\mu\nu\rho}^{4}$.

With the relation in Eq.~\eqref{eq: relation1-+}, we shall use the interpolating currents $J_{\mu}^{1}, J_{\mu}^{5}$ to investigate the $1^{-+}$ heavy quarkonium double-gluon hybrid mesons, while we employ the currents $J_{\mu\nu\rho}^{1}, J_{\mu\nu\rho}^{2}, J_{\mu\nu\rho}^{3}, J_{\mu\nu\rho}^{4}$ in Eq.~\eqref{eq:currentoperator3} to study the $2^{+-}$ hybrid mesons.

\section{QCD sum rules for the heavy quarkonium double-gluon hybrids}\label{sec3}
The two-point correlation function of the interpolating current is defined as
\begin{align}
\label{correlation1}
\Pi(p^2)=&i\int d^4x e^{ip\cdot x}\left\langle 0\left|T\left[J(x) J^\dagger(0)\right]\right|0\right\rangle\, .
\end{align}
At the hadron level, the correlation function can be described by the dispersion relation 
\begin{align}
\Pi\left(p^{2}\right)=&\frac{\left(p^{2}\right)^{N}}{\pi} \int_{4m_{Q}^{2}}^{\infty} \frac{\operatorname{Im} \Pi(s)}{s^{N}\left(s-p^{2}-i \epsilon\right)} d s\nonumber\\
&+\sum_{n=0}^{N-1} b_{n}\left(p^{2}\right)^{n}\, ,
\label{Cor-Spe}
\end{align}
where $b_n$ are the subtraction constants and will be removed by performing the Borel transform. The imaginary part of the correlation function is usually defined as the spectral function 
\begin{align}
    \rho(s)\equiv&\frac{1}{\pi}\text{Im} \Pi(s)\nonumber\\
    =&\sum_{n}\delta(s-m_n^2)\langle 0\lvert J\rvert n\rangle \langle n \lvert J^{\dagger} \rvert 0 \rangle \nonumber\\
    =&f_X^2\delta(s-m_X^2)+\cdots,\label{eq: rhos}
\end{align}
in which the ``one pole dominant narrow resonance'' approximation is adopted in the last step, and ``$\cdots$'' represents the contributions from the  continuum and higher excited states, and $f_X$, $m_X$ are the coupling constant and hadron mass of the lowest lying resonance respectively.

At the quark-gluon level, the correlation function can be calculated via the operator product expansion (OPE) method. We shall calculate the correlation functions at the leading order (LO) of $\alpha_s$ and consider the nonperturbative effects of condensates up to dimension eight. For the dimension-eight four-gluon condensate, we apply the factorization assumption to express it as $\left\langle \alpha_{s}GG\right\rangle^2$, which is a common approach in QCD sum rules for estimating the values of the high dimension condensates~\cite{Shifman:1978bx,Reinders1985}.

Here, we present the heavy quark propagator up to the one-gluon emission term in momentum space
\begin{align}
    iS_{ab}(p)=&\frac{i\delta_{ab}}{\cancel{p}-m_Q}\nonumber\\
    &-\frac{i}{4}g_{s}T^{r}_{ab}G^{r}_{\mu\nu}\frac{\sigma^{\mu\nu}(\cancel{p}+m_Q)+(\cancel{p}+m_Q)\sigma^{\mu\nu}}{(p^2-m_Q^2)^2}\,,\label{eq: quarkpropagator}
\end{align}
where $Q=c~\text{or}~b$. Similarly, the propagator for the gluon field strength $G^{r}_{\mu\nu}$ up to the one-gluon emission term in momentum space is given by
\begin{equation}\label{eq: gluon_propagator}
    \begin{aligned}
        &iS_{\alpha\mu,\, \beta\nu}^{rs}(p)\\
        =&\frac{-i\delta^{rs}}{p^2+i\epsilon}\bigg[g_{\mu\nu}p_{\alpha}p_{\beta}-g_{\mu\beta}p_{\alpha}p_{\nu}+(\mu\leftrightarrow \alpha,~\nu\leftrightarrow \beta)\bigg]\\
        &+\frac{i g_sf^{rst}}{(p^2+i\epsilon)^2}G^{t,\,\rho\sigma}\bigg[2 p_{\alpha}g_{\mu\rho}(p_{\beta}g_{\nu\sigma}-p_{\nu}g_{\beta\sigma})\\
        &+p_{\alpha}p_{\sigma}(g_{\mu\beta}g_{\nu\rho}-g_{\mu\nu}g_{\beta\rho})+(\mu\leftrightarrow \alpha,~\nu\leftrightarrow \beta)\bigg]\\
        &+\frac{i g_sf^{rst}}{2(p^2+i\epsilon)^2}G^{t,\,\rho\sigma}\bigg[2 g_{\alpha\sigma}p_{\rho}(g_{\mu\beta}p_{\nu}-g_{\mu\nu}p_{\beta})\\
        &-p^2g_{\alpha\sigma}(g_{\mu\beta}g_{\nu\rho}-g_{\mu\nu}g_{\beta\rho})+(\mu\leftrightarrow \alpha,~\nu\leftrightarrow \beta)\bigg]\,.
    \end{aligned}
\end{equation}
We have to point out that the one-gluon emission term in the expression of the propagator for the gluon field strength in Ref.~\cite{Tang:2021zti} was incorrect, indicating that the results regarding the tri-gluon condensate in Ref.~\cite{Tang:2021zti} were not reliable. 

\begin{figure}[ht!]
  \centering
  \subfigure[]{
      \includegraphics[width=0.1\textwidth]{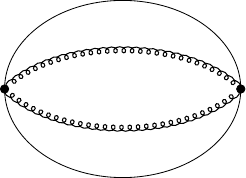}
      \label{fig: feynman_diagram_a}
      }\qquad
  \subfigure[]{
      \includegraphics[width=0.1\textwidth]{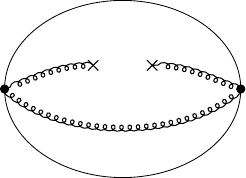}
      \label{fig: feynman_diagram_b}
      }\qquad
      \subfigure[]{
      \includegraphics[width=0.1\textwidth]{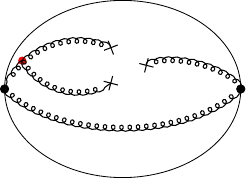}
      \label{fig: feynman_diagram_c}
      }\qquad
      \subfigure[]{
      \includegraphics[width=0.1\textwidth]{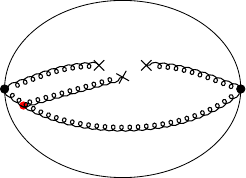}
      \label{fig: feynman_diagram_d}
      }\qquad
      \subfigure[]{
      \includegraphics[width=0.1\textwidth]{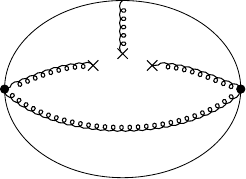}
      \label{fig: feynman_diagram_e}
      }\qquad
      \subfigure[]{
      \includegraphics[width=0.1\textwidth]{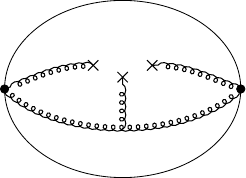}
      \label{fig: feynman_diagram_f}
      }\qquad
      \subfigure[]{
      \includegraphics[width=0.1\textwidth]{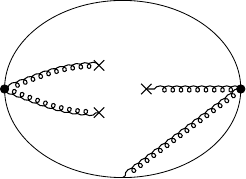}
      \label{fig: feynman_diagram_g}
      }\qquad
      \subfigure[]{
      \includegraphics[width=0.1\textwidth]{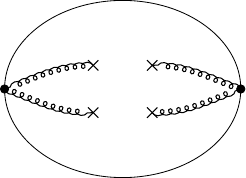}
      \label{fig: feynman_diagram_h}
      }\qquad
      \caption{The LO Feynman diagrams for the $\bar QGGQ$ hybrid meson systems up to dimension eight gluon condensate. Diagrams related by symmetry are not shown.}
      \label{fig: feynman_diagram}
\end{figure}

The Feynman diagrams considered up to dimension-eight condensates at the leading order of $\alpha_s$ are shown in Figure~\ref{fig: feynman_diagram}. We find that the diagram in Figure~\ref{fig: feynman_diagram_g} contributes to the correlation functions of the  currents with antisymmetric glueball operator $f^{rst}G^{s}_{\mu\nu}G^{t}_{\alpha\beta}$, but does not contribute to the currents with symmetric glueball operator $d^{rst}G^{s}_{\mu\nu}G^{t}_{\alpha\beta}$. This diagram is particularly important for $J_{\mu}^{1/2}$, in which the contribution from the dimension-4 condensate vanishes.

\begin{figure}[htbp]
    \centering
    \subfigure[]{
        \includegraphics[width=0.1\textwidth]{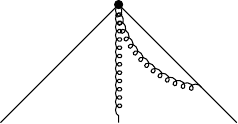}
        \label{fig: counter_diagram_a}
        }\qquad
    \subfigure[]{
        \includegraphics[width=0.1\textwidth]{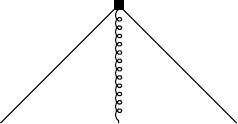}
        \label{fig: counter_diagram_b}
        }\qquad
    \subfigure[]{
        \includegraphics[width=0.1\textwidth]{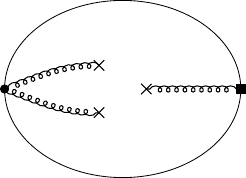}
        \label{fig: counter_diagram_c}
        }\qquad
    \caption{Diagram (a) is a subgraph of Figure~\ref{fig: feynman_diagram_g}, raising the non-local divergence exhibited in Figure~\ref{fig: feynman_diagram_g}. Diagram (b) with the black square represents the counter term vertex generated from the divergence of diagram (a). Diagram (c) with the counter term vertex will eliminate the non-local divergence in Figure~\ref{fig: feynman_diagram_g}.}
    \label{fig: counter_diagram}
\end{figure}

It is worth to mention that there exists non-local divergence in Figure~\ref{fig: feynman_diagram_g}, which cannot be simply removed by the subtraction scheme as shown in Eq.~\eqref{Cor-Spe}. To eliminate such non-local divergence in Figure~\ref{fig: feynman_diagram_g}, we employ the diagrammatic renormalization approach outlined in Refs.~\cite{Collins:1984xc,Muta:1998vi,deOliveira:2022eeq}. The renormalization process is illustrated in Figure~\ref{fig: counter_diagram}, where Figure~\ref{fig: counter_diagram_b} represents the counter term vertex introduced to cancel out the divergence originating from Figure~\ref{fig: counter_diagram_a}. The vertices of currents $J_{\mu}^{1}, J_{\mu\nu\rho}^{3}, J_{\mu\nu\rho}^{4}$, along with their corresponding counter terms and the associated renormalization coefficients are presented in Table~\ref{table: renormalization_constant}, in \ref{appendix: renormalization}. In next section, we will explore the importance of Figure~\ref{fig: feynman_diagram_g} in contributing to the tri-gluon condensate.

The Borel transform is applied to the correlation function at both the hadron and quark-gluon levels to enhance the convergence of OPE and reduce the impact of contributions from the continuum and higher excited states. Based on the hadron-quark duality, one can establish the QCD sum rules for the heavy quarkonium double-gluon hybrid  mesons 
\begin{align}
    \mathcal{L}_{k}(s_0,\,M_{B}^2)=&f_{X}^{2}(m_{X}^{2})^{k}e^{-m_X^2/M_B^2}\nonumber\\
    =&\int_{s_<}^{s_0}ds s^{k}\rho_{OPE}(s)e^{-s/M_B^2}\,,\label{eq: laplace_sr}
\end{align}
where $s_0$ is the continuum threshold and $M_B$ is the Borel mass. Then we can extract the lowest lying hadron mass
\begin{align}\label{eq: mass}
    m_X^2=\frac{\mathcal{L}_{1}(s_0,\,M_{B}^2)}{\mathcal{L}_{0}(s_0,\,M_{B}^2)}\,.
\end{align}

\section{Numerical analysis}\label{sec4}
To perform the QCD sum rule numerical analyses, we use the following values of the heavy quark masses, the strong coupling and the QCD condensates~\cite{ParticleDataGroup:2022pth,Narison:2011xe,Narison:2018dcr}:
\begin{equation}\label{eq: parameter_sr}
    \begin{aligned}
        m_{c}(\mu=\overline{m} _{c})&=(1.27\pm 0.02)~\text{GeV}\,,\\
        \alpha_{s}(\mu=\overline{m}_{c})&=0.38\pm 0.03\,,\\
        m_{b}(\mu=\overline{m}_{b})&=4.18^{+0.04}_{-0.03}~\text{GeV}\,,\\
        \alpha_{s}(\mu=\overline{m}_{b})&=0.223\pm 0.008\,,\\
        \GGb&=(6.35\pm 0.35)\times 10^{-2}~\text{GeV}^4\,,\\
        \GGGb&=(8.2\pm 1)\times \GGb~\text{GeV}^2\,,
    \end{aligned}
\end{equation}
where the heavy quark masses and strong coupling at the corresponding energy scales are obtained by using the two-loop QCD perturbation theory in the $\overline{\text{MS}}$ scheme.

 The hadron mass in Eq.~\eqref{eq: mass} depends on the continuum threshold $s_0$ and Borel mass $M_B$. The appropriate ranges of these two parameters are crucial for obtaining accurate sum rule predictions of the hadron mass. In QCD sum rules, the convergence of OPE and the pole contribution (PC) behaviors are typically analyzed to establish the optimal parameter working regions. To ensure satisfactory OPE convergence, it is necessary that the contributions of the condensates with $D=6$ and $D=8$ are kept below $20\%$ and $10\%$ respectively,
\begin{equation}\label{eq: convergence}
    \begin{aligned}
        R_{D=6}&=\left\lvert\frac{\mathcal{L}_{0}^{D=6}(\infty,M_B^2)}{\mathcal{L}_{0}^{total}(\infty,M_B^2)}\right\rvert<20\%\,,\\
        R_{D=8}&=\left\lvert\frac{\mathcal{L}_{0}^{D=8}(\infty,M_B^2)}{\mathcal{L}_{0}^{total}(\infty,M_B^2)}\right\rvert<10\%\, .
    \end{aligned}
\end{equation}
This requirement can give the lower bound on $M_B^2$. Additionally, we require that the pole contribution must exceed $40\%$ to determine the upper bound of $M_B^2$,
\begin{equation}\label{eq: PC}
    \text{PC}=\left\lvert\frac{\mathcal{L}_{0}^{total}(s_0,M_B^2)}{\mathcal{L}_{0}^{total}(\infty,M_B^2)}\right\rvert>40\%\,.
\end{equation}
With these two requirements, the Borel window of $M_B^2$ can be determined and an optimal value of the continuum threshold $s_{0}$ can also be obtained by minimizing the dependence of the hybrid mass $m_{X}$ on the Borel parameter $M_B^2$.

\begin{figure}[ht!]
  \centering
  \subfigure[]{
      \includegraphics[width=0.42\textwidth]{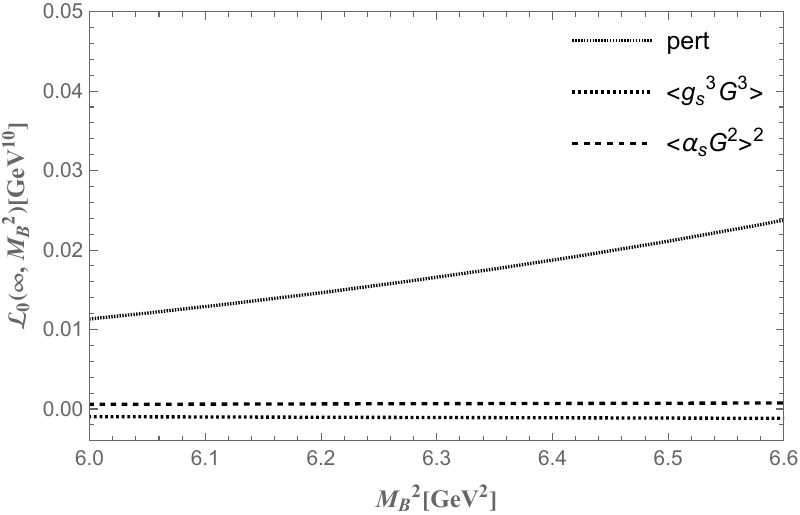}
      \label{fig: ope_J_mu1/2_a}
      }
  \subfigure[]{
      \includegraphics[width=0.42\textwidth]{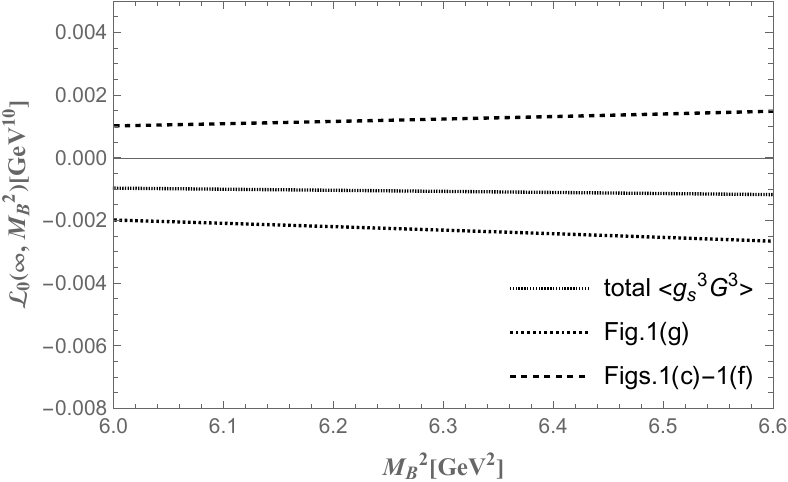}
      \label{fig: ope_J_mu1/2_b}
      }
  \caption{For the charmonium hybrid system with the interpolating current $J_{\mu}^{1}$ carrying $J^{PC}=1^{-+}$: (a) convergence of the OPE series; (b) important contribution from the Feynman diagram Figure~\ref{fig: feynman_diagram_g} to the tri-gluon condensate.}
  \label{fig: ope_J_mu1/2}
\end{figure}

In Figure~\ref{fig: ope_J_mu1/2}, we present the OPE series behavior for the  charmonium hybrid system using the current $J_{\mu}^{1}$, where the contribution of the dimension-4 gluon condensate vanishes. So the dominant non-perturbative effect comes from the dimension-6 tri-gluon condensate. Here, we highlight the significant contribution of the Feynman diagram shown in Figure~\ref{fig: feynman_diagram_g} to the tri-gluon condensate, a contribution that was overlooked in previous studies~\cite{Chen:2021smz,Su:2023jxb,Su:2023aif,Su:2022fqr,Tang:2021zti}. As shown in Figure~\ref{fig: ope_J_mu1/2_b}, the tri-gluon condensate contribution from Figure~\ref{fig: feynman_diagram_g} is negative whereas those from Figs.~\ref{fig: feynman_diagram_c}-\ref{fig: feynman_diagram_f} are positive. This leads to a negative value for the total contribution of tri-gluon condensate. Such a big improvement could potentially impact the stability of the sum rule and the  accuracy of the mass prediction. The analogous impact of the diagram in Figure~\ref{fig: feynman_diagram_g} on the tri-gluon condensate is also observed in the  currents $J_{\mu\nu\rho}^{3}$ and $J_{\mu\nu\rho}^{4}$, albeit with a lesser effect on the sum rule behaviors compared to the $J_{\mu}^{1}$ current.

\begin{figure}[ht!]
    \centering
    \subfigure[]{
        \includegraphics[width=0.42\textwidth]{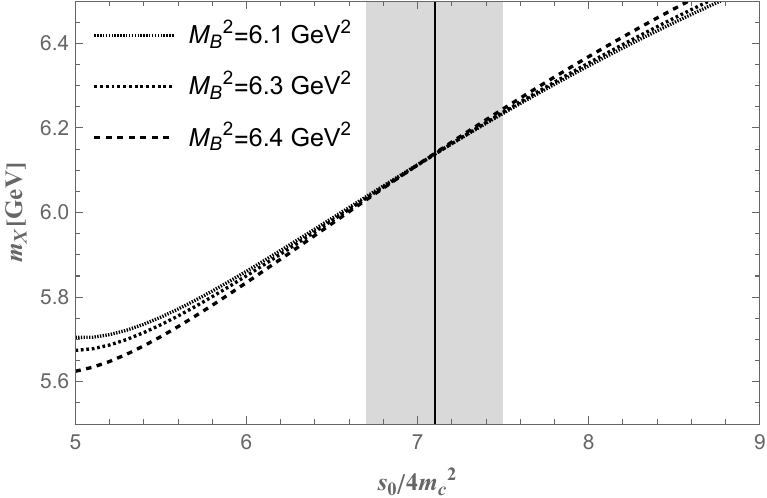}
        \label{fig: mass_J_mu1/2_a}
        }
    \subfigure[]{
        \includegraphics[width=0.42\textwidth]{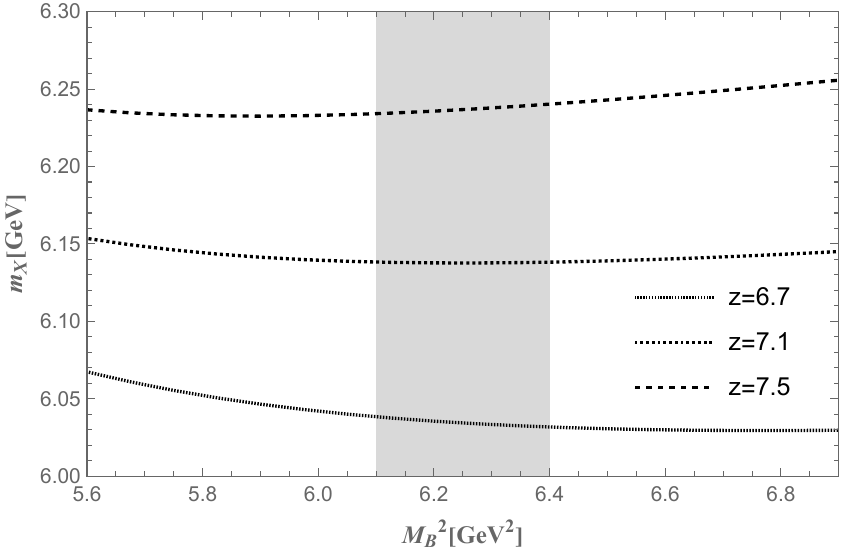}
        \label{fig: mass_J_mu1/2_b}
        }
    \caption{The variations of charmonium hybrid mass $m_X$ with respect to $z=s_0/4m_c^2$ and $M_B^2$ for $J_{\mu}^{1}$ with $J^{PC}=1^{-+}$.}
    \label{fig: mass_J_mu1/2}
  \end{figure}

For the charmonium hybrid system with current $J_{\mu}^{1}$ carrying $J^{PC}=1^{-+}$, we determine the parameter working regions as $6.12$ GeV$^2\leq M_B^2\leq 6.42$ GeV$^2$ and $z=s_0/4m_c^2=7.1\pm 0.4$, after the investigation of the OPE convergence, pole contribution and mass sum rule stability criteria. In Figure~\ref{fig: mass_J_mu1/2}, we show the changes in the charmonium hybrid mass in relation to the variables $z=s_{0}/4m_{c}^2$ and $M_{B}^2$. The mass sum rules exhibit high stability within the Borel window (depicted in the gray region) in Figure~\ref{fig: mass_J_mu1/2_b}. This enables the prediction of the hybrid mass as
\begin{equation}
m_X=6.14\pm0.19 ~\text{GeV}\, .
\end{equation}

After performing the QCD sum rule analyses for all interpolating currents, we present the numerical results for the charmonium and bottomonium hybrids in Table \ref{table: mass_charmonium} and \ref{table: mass_bottomonium}, respectively. The uncertainties from the heavy quark masses, the strong coupling constants and the condensates are taken into account. 
One can observe that the pole contributions for the currents  $J_{\mu\nu\rho}^1$ and $J_{\mu\nu\rho}^3$ with $J^{PC}=2^{+-}$ are notably smaller compared to those of other currents, indicating the weak couplings of these two currents to the heavy quarkonium double-gluon hybrid states. Nevertheless, we still present the mass predictions for these two currents, which are much lower than masses extracted from $J_{\mu\nu\rho}^2$ and $J_{\mu\nu\rho}^4$ with the same quantum numbers. 

\begin{table*}[t!] 
    \centering
    \renewcommand\arraystretch{1.8}
    \caption{Numerical results for the $\bar cGGc$ charmonium double-gluon hybrid states. In the third column, ``A'' and ``S'' represent the antisymmetric and symmetric glueball operators $f^{rst}G^{s}_{\mu\nu}G^{t}_{\alpha\beta}$ and $d^{rst}G^{s}_{\mu\nu}G^{t}_{\alpha\beta}$ in the interpolating currents, respectively. }
    \setlength{\tabcolsep}{2mm}
    {
        \begin{tabular}{ccccccc}
            \hline\hline
        Currents  & $J^{PC}$ & Glueball Operator & $s_{0}/4m_c^2$ & $M_B^2~(\text{GeV}^2)$ & $m_{X}~(\text{GeV})$ & PC\\ \hline
            $J_{\mu}^{1}$ & $1^{-+}$ & A & $7.10\pm 0.40$ & $6.12 - 6.42$ & $6.14\pm 0.19$ & $45.9\%$\\
            $J_{\mu}^{5}$ & $1^{-+}$ & S & $9.70\pm 0.52$ & $8.15 - 8.56$ & $7.21\pm 0.14$ & $46.8\%$\\
            $J_{\mu\nu\rho}^{1}$ & $2^{+-}$ & S & $3.24\pm 0.53$ & $7.75 - 8.75$ & $4.22\pm 0.18$ & $2.2\%$\\
            $J_{\mu\nu\rho}^{2}$ & $2^{+-}$ & S & $7.71\pm 0.44$ & $6.77 - 7.25$ & $6.41\pm 0.17$ & $46.6\%$\\
            $J_{\mu\nu\rho}^{3}$ & $2^{+-}$ & A & $3.74\pm 0.28$ & $5.00 - 5.50$ & $4.55\pm 0.17$ & $12\%$\\
            $J_{\mu\nu\rho}^{4}$ & $2^{+-}$ & A & $7.40\pm 0.37$ & $6.16 - 6.51$ & $6.33\pm 0.17$ & $44.3\%$\\
            \hline\hline
        \end{tabular}
    }
    
\label{table: mass_charmonium}
\end{table*}
\begin{table*}[t!] 
    \centering
    \renewcommand\arraystretch{1.8}
    \caption{Numerical results for the $\bar bGGb$ bottomonium double-gluon hybrid states.}
    \setlength{\tabcolsep}{2mm}
    {
        \begin{tabular}{ccccccc}
            \hline\hline
        Currents  & $J^{PC}$ & Glueball Operator & $s_{0}/4m_b^2$ & $M_B^2~(\text{GeV}^2)$ & $m_{X}~(\text{GeV})$ & PC\\ \hline
            $J_{\mu}^{1}$ & $1^{-+}$ & A & $3.38\pm 0.09$ & $23.55 - 24.43$ & $14.26\pm 0.19$ & $45.8\%$\\
            $J_{\mu}^{5}$ & $1^{-+}$ & S & $3.04\pm 0.12$ & $18.13 - 19.00$ & $13.71\pm 0.20$ & $44.1\%$\\
            $J_{\mu\nu\rho}^{1}$ & $2^{+-}$ & S & $2.16\pm 0.08$ & $15.31 - 15.92$ & $11.67\pm 0.26$ & $19\%$\\
            $J_{\mu\nu\rho}^{2}$ & $2^{+-}$ & S & $2.58\pm 0.11$ & $14.76 - 15.44$ & $12.58\pm 0.16$ & $49.8\%$\\
            $J_{\mu\nu\rho}^{3}$ & $2^{+-}$ & A & $1.52\pm 0.16$ & $12.85 - 14.65$ & $9.85\pm 0.43$ & $5.3\%$\\
            $J_{\mu\nu\rho}^{4}$ & $2^{+-}$ & A & $2.88\pm 0.13$ & $18.20 - 19.10$ & $13.31\pm 0.19$ & $43.3\%$\\
            \hline\hline
        \end{tabular}
    }

\label{table: mass_bottomonium}
\end{table*}

\section{Summary and discussion}\label{sec5}
In this paper, we discussed the color structure of the double-gluon hybrid state and considered the $\bar QGGQ$ operators in the octet-octet $(\mathbf{8}_{[\bar{Q}Q]} \otimes \mathbf{8}_{[GG]})$ color structure to construct currents with $J^{PC}=1^{-+}$ and $2^{+-}$. We calculated the two-point correlation functions and spectral functions for the heavy quark systems up to dimension-8 condensate, with the corresponding Feynman diagrams depicted in Figure~\ref{fig: feynman_diagram}. We noted that the diagram in Figure~\ref{fig: feynman_diagram_g} was neglected in prior studies, and the diagram could potentially contribute significantly to the OPE series of the currents involving the antisymmetric glueball operator $f^{rst}G^{s}_{\mu\nu}G^{t}_{\alpha\beta}$. 

The QCD sum rule analyses are employed for both the double-gluon charmonium and bottomonium hybrid mesons with $J^{PC}=1^{-+}$ and $2^{+-}$, and the numerical results are collected in Table \ref{table: mass_charmonium} and \ref{table: mass_bottomonium}, respectively. The pole contributions of the currents $J_{\mu\nu\rho}^1$ and $J_{\mu\nu\rho}^3$ are pretty small, indicating the weak couplings of these two currents to the heavy quarkonium double-gluon hybrid states. We also observe that the charmonium $\bar cGGc$ hybrid mesons with symmetric glueball operators are heavier than those with antisymmetric glueball operators in both the $1^{-+}$ and $2^{+-}$ channels. However, the situation is reversed for the bottomonium $\bar bGGb$ system, in which the hybrid mesons with symmetric glueball operators are lighter than those with antisymmetric glueball operators. 

\begin{figure}[hb!]
    \centering
    \subfigure[]{
        \includegraphics[width=0.42\textwidth]{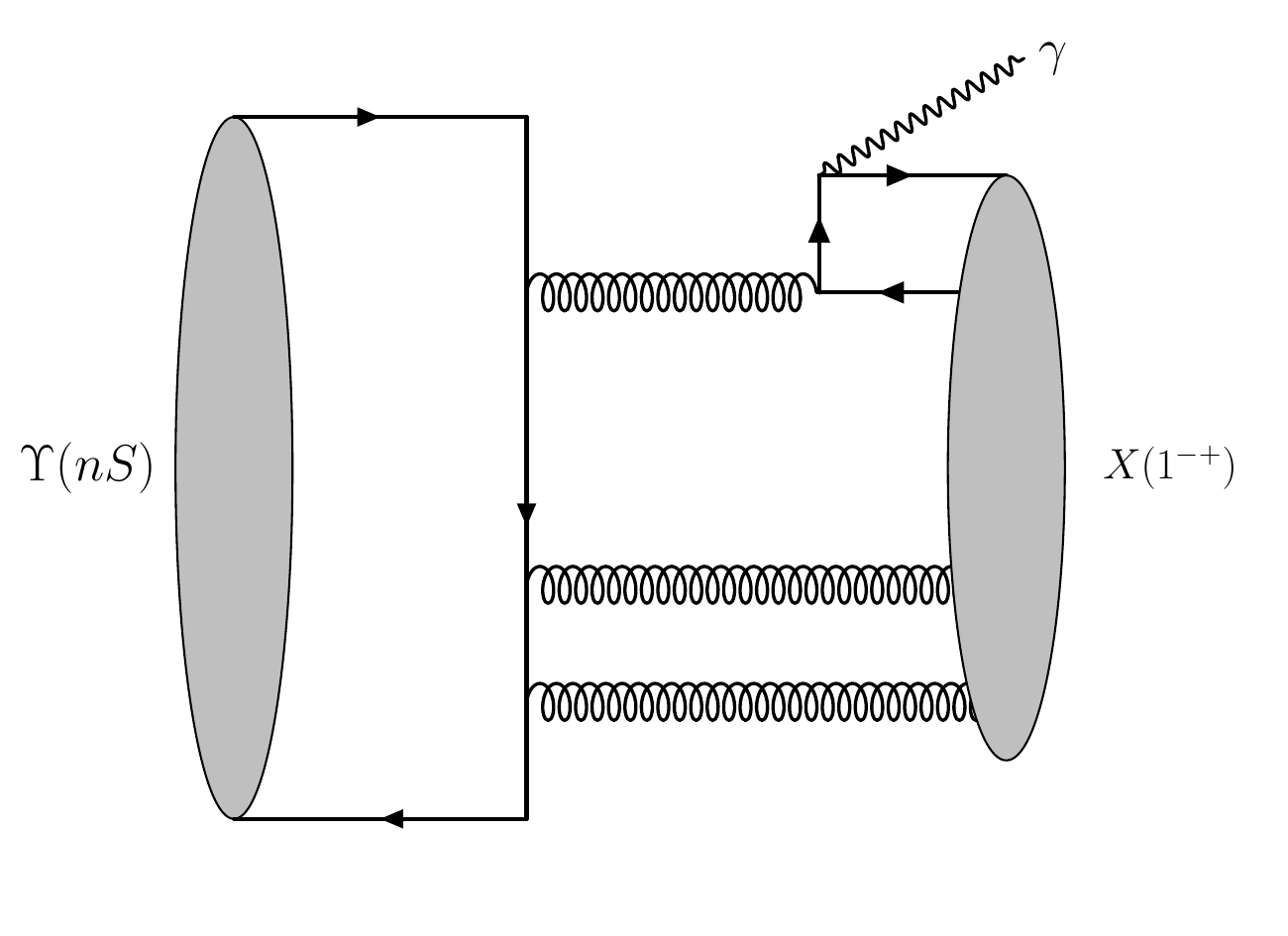}
        \label{fig: production_charmonium_a}
        }\qquad
    \subfigure[]{
        \includegraphics[width=0.42\textwidth]{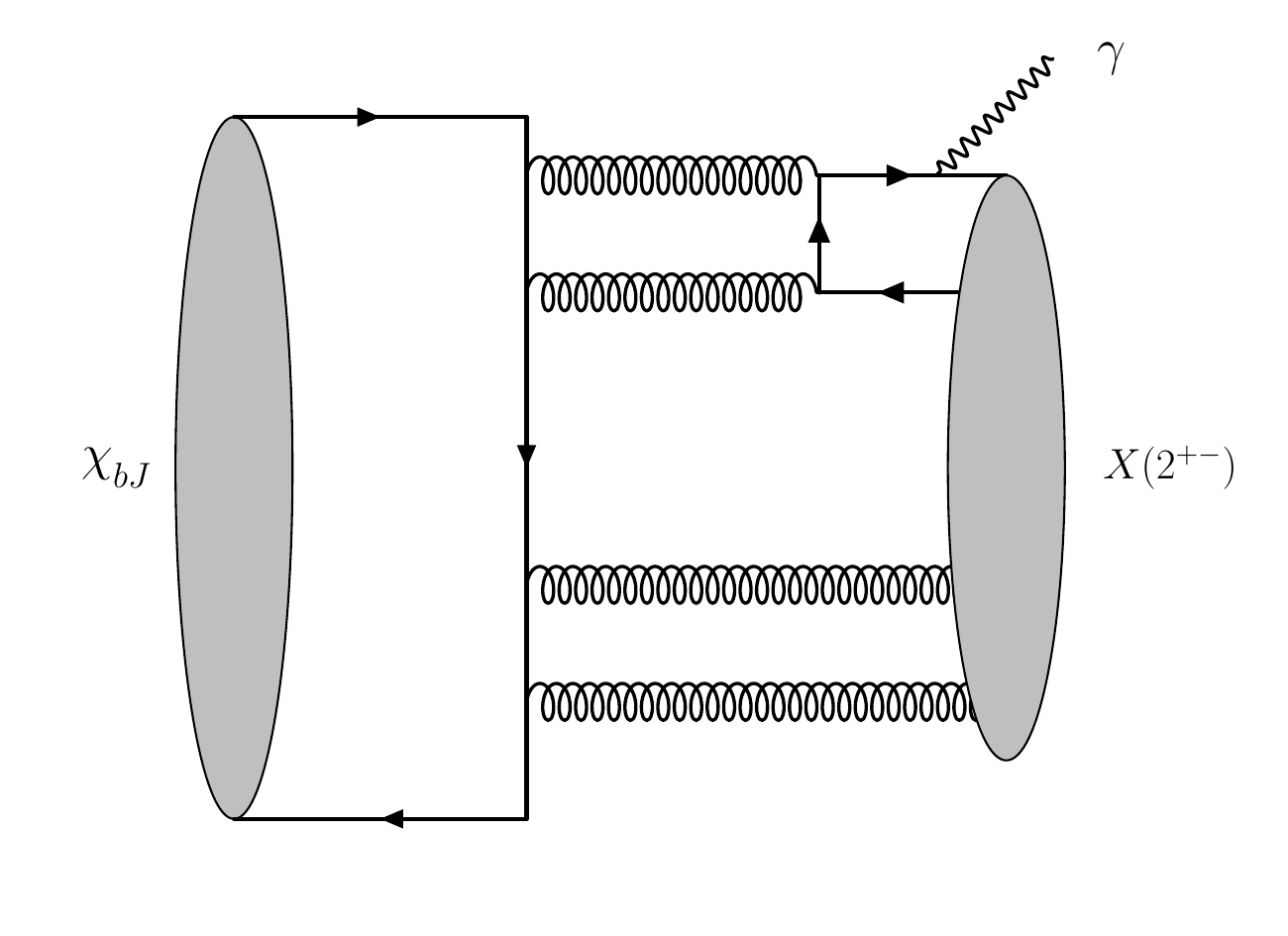}
        \label{fig: production_charmonium_b}
        }\qquad
    \caption{The possible production mechanisms of the charmonium double-gluon hybrids with $J^{PC}=1^{-+}$ and $2^{+-}$.}
    \label{fig: production_charmonium}
\end{figure}

The charmonium double-gluon hybrid mesons can decay into the final states with two or three mesons, as well as two baryons. For more detailed discussions, one can refer to Ref.~\cite{Su:2023aif}. The possible production mechanisms of the charmonium double-gluon hybrids with $J^{PC}=1^{-+}$ and $2^{+-}$ are similar to the production of $\eta_{1}(1855)$ in the $J/\psi$ radiative decay process~\cite{BESIII:2022riz,BESIII:2022iwi,Chen:2022isv}. They could be generated through the three-gluon and four-gluon emission processes in the radiative decay of $\Upsilon(n S)$ and $\chi_{bJ}$ respectively, as depicted in Figure~\ref{fig: production_charmonium}. With the abundance of bottomonium mesons available in BelleII experiment, there is potential to detect these charmonium double-gluon hybrids in the future. Further explorations of these hybrid mesons using a variety of theoretical and phenomenological methods are both beneficial and anticipated.

\section*{Acknowledgments}
This work is supported by the National Natural Science Foundation of China under Grant No. 12175318 and No. 12305147,  the Natural Science Foundation of Guangdong Province of China under Grant No. 2022A1515011922, the National Key R$\&$D Program of China under Contracts No. 2020YFA0406400.

\appendix
\section{Relations of the operators}\label{appendix: relation}
We know that $\widetilde{G}^{s}_{\mu\nu}=\frac{1}{2}\varepsilon _{\mu\nu\alpha\beta}G^{s,\,\alpha\beta}$ and $G^{s}_{\mu\nu}=-\frac{1}{2}\varepsilon _{\mu\nu\alpha\beta}\widetilde{G}^{s,\,\alpha\beta}$, so we find two relations as below
\begin{equation}\label{eq: relationglueball}
    \begin{aligned}
        \widetilde{G}^{s}_{\mu\nu}G^{t,\,\nu\rho}&=-\frac{1}{4}\varepsilon_{\mu\nu\gamma\lambda}\varepsilon^{\nu\rho\alpha\beta}G^{s,\,\gamma\lambda}\widetilde{G}^{t}_{\alpha\beta}\\
        &=-\frac{1}{4}(4G^{s,\,\nu\rho}\widetilde{G}^{t}_{\mu\nu}+2\delta^{\,\,\rho}_{\mu}G^{s,\,\alpha\beta}\widetilde{G}^{t}_{\alpha\beta})\\
        &=-G^{s,\,\nu\rho}\widetilde{G}^{t}_{\mu\nu}-\frac{1}{2}\delta^{\,\,\rho}_{\mu}G^{s,\,\alpha\beta}\widetilde{G}^{t}_{\alpha\beta}\,,\\
        \widetilde{G}^{s}_{\mu\nu}\widetilde{G}^{t,\,\nu\rho}&=\frac{1}{4}\varepsilon_{\mu\nu\gamma\lambda}\varepsilon^{\nu\rho\alpha\beta}G^{s,\,\gamma\lambda}G^{t}_{\alpha\beta}\\
        &=\frac{1}{4}(4G^{s,\,\nu\rho}G^{t}_{\mu\nu}+2\delta^{\,\,\rho}_{\mu}G^{s,\,\alpha\beta}G^{t}_{\alpha\beta})\\
        &=G^{s,\,\nu\rho}G^{t}_{\mu\nu}+\frac{1}{2}\delta^{\,\,\rho}_{\mu}G^{s,\,\alpha\beta}G^{t}_{\alpha\beta}\,,
    \end{aligned}
\end{equation}
in which we have performed the well-known identity
\begin{align}\label{eq: levicivitaidentity}
    \varepsilon_{\mu\nu\gamma\lambda}\varepsilon^{\nu\rho\alpha\beta}=
    \begin{vmatrix}
        \delta^\rho_\mu&\delta^\rho_\gamma&\delta^\rho_\lambda\\
        \delta^\alpha_\mu&\delta^\alpha_\gamma&\delta^\alpha_\lambda\\
        \delta^\beta_\mu&\delta^\beta_\gamma&\delta^\beta_\lambda
    \end{vmatrix}.
\end{align}
By using the results in Eq.~\eqref{eq: relationglueball}, we obtain
\begin{subequations}\label{eq: glueballrelation}
    \begin{align}
     f^{rst}G^{s}_{\alpha\beta}G^{t,\,\alpha\beta}&=0\,,\label{eq: glueballrelation1} \\
        f^{rst}\widetilde{G}^{s}_{\mu\nu}\widetilde{G}^{t,\,\nu\rho}&=-f^{rst}G^{s}_{\mu\nu}G^{t,\,\nu\rho}\,,\label{eq: glueballrelation2}\\
      d^{rst}\widetilde{G}^{s}_{\mu\nu}G^{t,\,\nu\rho}&=d^{rst}G^{s}_{\mu\nu}\widetilde{G}^{t,\,\nu\rho}\nonumber\\
      &=-\frac{1}{4}\delta^{\,\,\rho}_{\mu}d^{rst}G^{s,\,\alpha\beta}\widetilde{G}^{t}_{\alpha\beta}\,.\label{eq: glueballrelation3}
        \end{align}
\end{subequations}
So the six currents in Eqs.~\eqref{eq: currentoperator1} are not independent 
\begin{align}\label{eq: relation1-+}
    J_{\mu}^{1}=-J_{\mu}^{2},\quad J_{\mu}^{3}=0,\quad J_{\mu}^{5}=J_{\mu}^{6}=-\frac{1}{4}J_{\mu}^{4}.
\end{align}
\section{Operator renormalization}\label{appendix: renormalization}
The renormalization is performed for the interpolating currents $J_{\mu}^{1}$, $J_{\mu\nu\rho}^{3}$ and $J_{\mu\nu\rho}^{4}$ with the antisymmetric glueball operator $f^{rst}G^{s}_{\mu\nu}G^{t}_{\alpha\beta}$, for which the Feynman diagram in Figure~\ref{fig: feynman_diagram_g} will give an important contribution to the tri-gluon condensate. Typically, the diagram in Figure~\ref{fig: counter_diagram_a} for the double-gluon hybrid operator has two kinds of vertices. For $J_{\mu}^{1}$, its vertices are $A_{J_{\mu}^{1}}=g_{s}^{2}f^{rst}G^{s}_{\mu\nu}T^{r}\gamma_{\rho}$ and $B_{J_{\mu}^{1}}=g_{s}^{2}f^{rst}G^{t,\,\nu\rho}T^{r}\gamma_{\rho}$. For $J_{\mu\nu\rho}^{3}$, its vertices are 
$A_{J_{\mu\nu\rho}^{3}}=g_{s}^{2}f^{rst}G^{s}_{\mu\nu}T^{r}\gamma_{\alpha}\gamma_{5}$ and $B_{J_{\mu\nu\rho}^{3}}=g_{s}^{2}f^{rst}G^{t,\,\alpha}_{\quad\rho}T^{r}\gamma_{\alpha}\gamma_{5}$. For $J_{\mu\nu\rho}^{4}$, its vertices are $A_{J_{\mu\nu\rho}^{4}}=g_{s}^{2}f^{rst}G^{s}_{\mu\nu}T^{r}\gamma_{\alpha}\gamma_{5}$ and $B_{J_{\mu\nu\rho}^{4}}=g_{s}^{2}f^{rst}\widetilde{G}^{t,\,\alpha}_{\quad\rho}T^{r}\gamma_{\alpha}\gamma_{5}$. It is clear that $A_{J_{\mu\nu\rho}^{3}}=A_{J_{\mu\nu\rho}^{4}}$ and $\widetilde{B}_{J_{\mu\nu\rho}^{3}}=B_{J_{\mu\nu\rho}^{4}}$, where $\widetilde{B}$ denotes the duality of $B$. We employ the $\overline{\text{MS}}$ scheme to evaluate the renormalization coefficients of the counter terms in Figure~\ref{fig: counter_diagram_b} using the dimensional regularization. The results are shown in Table~\ref{table: renormalization_constant}.
\begin{table*}[hb!]
    \centering
    \caption{The vertices and renormalization coefficients of counter terms for the diagram Figure~\ref{fig: counter_diagram_b}. $k_1$ and $k_2$ represent the  momenta of incoming and outgoing fermion lines, respectively. An overall factor $\frac{g_s}{16\pi^2}\left(\frac{1}{\epsilon}-\gamma_{E}+\log(4\pi)\right)$ for the renormalization coefficients is understood in the table. }
    \setlength{\tabcolsep}{2mm}{
        \renewcommand\arraystretch{1.38}
        \begin{tabular}{ccc}
            \hline\hline
            Vertices of currents & Vertices of counter terms& Renormalization coefficients\\
            \hline
            \multirow{7}{*}{$A_{J_{\mu}^{1}}$} & $i g_s^2 m_Q^2 T^{t}\left(\sigma_{\mu\nu}\gamma_{\rho}-\gamma_{\rho}\sigma_{\mu\nu}\right)$ &-$\frac{3}{4}$\\
            &$g_s^2 m_Q T^{t}\left(k_{1,\,\mu}\gamma_{\rho}\gamma_{\nu}-k_{1,\,\nu}\gamma_{\rho}\gamma_{\mu}\right)$ &$ \frac{3}{4}$\\
            &$i g_s^2 k_1^2 T^{t} \gamma_{\rho}\sigma_{\mu\nu}$ &$-\frac{1}{4}$\\
            &$g_s^2 T^{t}\gamma_{\rho} \left(k_{1,\,\mu}\cancel{k}_{1}\gamma_{\nu}-k_{1,\,\nu}\cancel{k}_{1}\gamma_{\mu}\right)$ & $\frac{1}{4}$\\
            & $g_s^2 m_Q T^{t}\left(k_{2,\,\mu}\gamma_{\nu}\gamma_{\rho}-k_{2,\,\nu}\gamma_{\mu}\gamma_{\rho}\right)$ & $\frac{3}{4}$\\
            &$i g_s^2 k_2^2 T^{t} \sigma_{\mu\nu}\gamma_{\rho}$& $\frac{1}{4}$\\
            &$g_s^2 T^{t}\left(k_{2,\,\mu}\gamma_{\nu}\cancel{k}_{2}-k_{2,\,\nu}\gamma_{\mu}\cancel{k}_{2}\right)\gamma_{\rho}$ &$\frac{1}{4}$\\
            \hline
            \multirow{5}{*}{$B_{J_{\mu}^{1}}$}& $g_s^2 m_Q^2 T^{s} \gamma_{\rho}$ &$-\frac{9}{2}$\\
            &$g_s^2 m_Q T^{s}\left(\cancel{k}_1\gamma_{\rho}+\gamma_{\rho}\cancel{k}_2\right)$ &$\frac{3}{4}$\\
            &$g_s^2 m_Q T^{s}(k_{1,\,\rho}+k_{2,\,\rho})$& $-3$\\
            &$g_s^2 T^{s} \gamma_{\rho}(k_1^2+k_2^2)$ & $1$\\
            &$g_s^2 T^{s}(k_{1,\,\rho}\cancel{k}_{1}+k_{2,\,\rho}\cancel{k}_{2})$ & $\frac{1}{2}$\\
            \hline
            \multirow{7}{*}{$A_{J_{\mu\nu\rho}^{3}}$ (or $A_{J_{\mu\nu\rho}^{4}}$)} & $i g_s^2m_Q^2T^{t}\left(\sigma_{\mu\nu}\gamma_{\alpha}\gamma_5-\gamma_{\alpha}\gamma_5\sigma_{\mu\nu}\right)$&$-\frac{3}{4}$ \\
            &$g_s^2m_QT^{t}\left(k_{1,\,\mu}\gamma_{\alpha}\gamma_5\gamma_{\nu}-k_{1,\,\nu}\gamma_{\alpha}\gamma_5\gamma_{\mu}\right)$ & $\frac{3}{4}$ \\
            &$i g_s^2k_1^2T^{t}\gamma_{\alpha}\gamma_5\sigma_{\mu\nu}$ &$-\frac{1}{4}$\\
            & $g_s^2T^{t}\gamma_{\alpha}\gamma_5\left(k_{1,\,\mu}\cancel{k}_{1}\gamma_{\nu}-k_{1,\,\nu}\cancel{k}_{1}\gamma_{\mu}\right) $& $\frac{1}{4}$\\
            &$g_s^2m_QT^{t}\left(k_{2,\,\mu}\gamma_{\nu}\gamma_{\alpha}\gamma_5-k_{2,\,\nu}\gamma_{\mu}\gamma_{\alpha}\gamma_5\right)$& $\frac{3}{4}$\\
            &$i g_s^2k_2^2T^{t}\sigma_{\mu\nu}\gamma_{\alpha}\gamma_5$ & $\frac{1}{4}$\\
            &$g_s^2T^{t}\left(k_{2,\,\mu}\cancel{k}_{2}\gamma_{\nu}-k_{2,\,\nu}\cancel{k}_{2}\gamma_{\mu}\right)\gamma_{\alpha}\gamma_5$&$\frac{1}{4}$\\
            \hline
            \multirow{5}{*}{$B_{J_{\mu\nu\rho}^{3}}$} & $g_s^{2}m_Q^2 T^{s} \gamma_{\rho}\gamma_5$ & $\frac{9}{2}$\\
            & $ g_s^{2}m_Q T^{s}\left(\cancel{k}_{1}\gamma_{\rho}\gamma_5+\gamma_{\rho}\gamma_5\cancel{k}_{2}\right)$ & $\frac{3}{4}$\\
            & $g_s^{2}m_Q T^{s}\gamma_5\left(k_{1,\,\rho}-k_{2,\rho}\right)$ & $-3$\\
            & $g_s^{2} T^{s} \left(k_{1}^2+k_{2}^2\right)\gamma_{\rho}\gamma_5$ &$-1$\\
            &$g_s^{2} T^{s}\left(k_{1,\,\rho}\cancel{k}_{1}+k_{2,\,\rho}\cancel{k}_{2}\right)\gamma_5$&$-\frac{1}{2}$\\
            \hline
            \multirow{3}{*}{$B_{J_{\mu\nu\rho}^{4}}$} & $g_s^{2}m_Q T^{s} \left(\sigma_{\rho\alpha}k_{1}^{\alpha}-\sigma_{\rho\alpha}k_{2}^{\alpha}\right)$ & $-\frac{3}{2}$\\
            & $i g_s^{2}T^{s}\left(k_{1}^2-k_{2}^2\right)\gamma_{\rho}$ & $\frac{5}{4}$ \\
            & $i g_s^{2}T^{s}\left(k_{1,\,\rho}\cancel{k}_{1}-k_{2,\rho}\cancel{k}_{2}\right)$ &$-\frac{1}{2}$\\
            \hline\hline
        \end{tabular}
    }
    
    \label{table: renormalization_constant}
\end{table*}

\end{document}